\documentclass[preprint,12pt]{elsarticle}

\usepackage{amssymb}

\journal{arXiv }
\date{2010/10/31}
\begin{document}
\pdfoutput = 1
\begin{frontmatter}


 \ead{lfjsde@gmail.com }

\title{Golden and Alternating, fast simple $O(\lg n)$  algorithms for Fibonacci}


\author{L. F. Johnson}

\address{University of Waterloo}

\begin{abstract}
Two very fast and simple  $O(\lg n)$ iteration algorithms for individual Fibonacci  numbers 
are given and compared to competing algorithms. A simple $O(\lg n)$ recursion is derived that 
can also be applied to Lucas. 
A formula is given to estimate the largest $n$, 
where $F_{n}$ does not overflow  the implementation's data type. The danger of timing runs on 
input that is too large for the computer representation leads to false research results.

\end{abstract}


\end{frontmatter}




\section{Introduction}
The determination of individual Fibonacci numbers, first defined in 1202, 
has an  interesting and extensive  literature [1,2]. 
The recursive definition  $\ F_{n}= F_{n-1} +F_{n-2}$ leads naturally 
to iteration. 
Defining $\ F_{0}$ as 0 incorporates two common statements of initial conditions 
by  starting  at $\ F_{0}$ or $\ F_{1}$ and  this fixes the 
value of $\ F_{n}$, which varies with different authors. Lucas numbers are 
generated by using initial conditions 1,3 and it is convenient to define $L_{0}=2$. [6]

There are a number of direct iterative solutions for this 
recursive definition that are $O(n)$ [6].
DeMoivre published a closed formula in 1730 that requires $n$ multiplications [1].    
$F_{n}$ was first shown to be found in $O(\lg n)$ time in 1978 [3] followed by 
improved algorithms [4,5], but the best method [5] is complicated 
 and only becomes theoretically effective for extremely large $n$.
 
The Fibonacci recursion formula can have other initial values.
It is easy to show by induction that for 
${\mathcal L}_{n} = {\mathcal L}_{n-1}+ {\mathcal L}_{n-2}$ ,
where ${\mathcal L}_{0}$ and ${\mathcal L}_{1}$ are non-negative integers,
  ${\mathcal L}_{n} =  {\mathcal L}_{1} * F_{n} + {\mathcal L}_{0}* F_{n-1}$. So all numbers 
${\mathcal L}_{n}$ can be found in $O(\lg n )$ time by these methods.
 
An experimental  comparison of algorithms in [4] did not include the algorithms in this paper, 
but did include the closed form DeMoivre formula as a {\it linear method} and  derived from 
it a $O(\lg n)$ recursive method they called the {\it Binet} algorithm: 
$F_{n}= \lceil F_{n/2}^2 \times \surd 5 \rceil$,
when $n= 2^{m}$. Because the derivation assumes $n= 2^{m}$, {\it Binet} does 
not work exactly for all $n$. 
This is a nice and interesting recursive result but only exact for  a subset of $ n$ that when  when $n=2^m$ , which uses two multiplications and 
the ceiling function in each call.

$$
\begin{array}{l}

fib(n)                                                                \\
\ \  if  \ n = 1   \ or \ n=2   \  return \  1                                            \\
 \ \  else \ return \  \lceil   (fib(n/2))^2 \sqrt 5 \  \rceil  \\                               \\

([5]  Fig2 \ Recursive \  Binet \ approximation\  Cull \  \&  \ Holloway )
\end{array}  
$$

 Using Lucas sequences [3], they [4] also constructed  an efficient 
$O(\lg n)$ algorithm, named here {\it Cullhow}, requiring only two multiplications 
in a loop step; this they claimed to be the best in theoretical and experimental 
comparisons of their algorithms. As expressed, {\it Cullhow} assumed $n= 2^{m}$, but 
the authors were aware of how to extend it to all $n$, and this was done by an other author in [5].
This algorithm is our base for comparsion and is given in 1.1.
These comparisons included the size of $F_{n}$, which  grows as $O(F_{n})$; thus, this 
execution cost is much higher than the simple iteration cost (which assumes that the 
cost of arithmetic is constant.) However, the number of bits 
required for computer representation is only $ \lfloor \lg F_{n} +1$, so the 
computer execution cost of arithmetic is actually based on $O(\lg  F_{n})$.

\subsection{Comparsion Algorithm by Takakaski  [5]}

$$
\begin{array}{l}

fib(n)                                                                \\
\ \  if  \ n = 0  \  return \  0                                            \\
 \ \  else \ if \  n = 1  \ return \ 1                               \\
\ \  else \  if  \ n = 2  \  return \ 1     \\
 \ \   else \\ 
   \   \ \  f  \leftarrow 1 \\
    \  \ \   l   \leftarrow  1 \\
\ \ \   sign \leftarrow -1 \\
\ \ \    mask \leftarrow 2^{\lfloor \log_2 n - 1}   \\
\ \ \    for\  i = 1  \ to  \ \lfloor \log_2 n-1 \\
\ \ \ \ \  temp \leftarrow f * f     \\
\ \ \ \  \ f \leftarrow (f+l)/2  \\
\ \ \ \ \  f \leftarrow 2 * (f*f) - 3*temp - 2*sign \\ 
\ \ \ \ \  l \leftarrow 5 * temp + 2* sign  \\ 
\ \ \ \ \  sign \leftarrow 1  \\
\ \ \ \ \   if( n \  \& \  mask) \neq  0 \\
\ \ \ \ \ \  temp \leftarrow f   \\
\ \ \ \ \ \  f \leftarrow (f+l)/2 \\
\ \ \ \ \ \  l \leftarrow f+2*temp  \\
\ \ \ \ \ \  sign \leftarrow -1  \\
\ \  mask \leftarrow mask /2  \\
if(n \ \& \ mask)=0  \\
\  f \leftarrow f*l  \\
else  \\
\  f \leftarrow (f+l)/2   \\
\   f \leftarrow f*l - sign  \\
return \ f  \\  \\

([5]  Fig3 \ Presented \ product \ of  \ Lucas  \ numbers \ to \ compute \ F_n \ for \ arbitrary \ n )
\end{array}  
$$

\section{ Algorithms}
We present two effective representations of applying the DeMoive formula 
that have not appeared in other papers and give a new integer algorithm that uses only 
two multiplications in a novel way, which does not require the use of Lucas numbers [5]. 
\subsection{ Golden,  a real number iteration process, and Rgolden, its recursion} 
The DeMoive closed form $F_n = {\phi^n - \overline  \phi^n \over \surd 5}$ can be replaced 
by $F_n = \lceil ( {\phi^n \over \surd 5}- 0.5)$ [1]. 
Based on {\it efficiently computed powers} of the 
golden ratio, $\phi = {( 1 + \surd 5) \over 2}$, the Algorithm 
{\it Golden} is not only $O(\lg n) \ \forall n \geq 0 $  but uses fewer
 multiplications than other methods. It can also find Lucas numbers directly 
in $O(\lg n)$ time 
because $L_n = {\phi^n - \overline  \phi^n }$, 
where $ \overline \phi  = {( 1 - \surd 5) \over 2}$ . In the following discusion, 
we assume that $\phi$ exists as a constant just as does $\pi$.
$$
\begin{array}{l}
 Algorithm: \ Golden(n) \ given \ n \geq 0, return \ F_{n}   \\
   
        gi \leftarrow   \phi                    \\     
        
    if ( odd(n)) \ then \ F \leftarrow \phi  \  else \ F \leftarrow 1     \\
    i \leftarrow n                                 \\
    \  \ \ while (i>1)                               \\

    \ \ \ \{  \ i=i/2; \ gi \leftarrow gi * gi; \ if ( odd(i)) F \leftarrow gi*F  \}           \\
          
    return \  \lceil( \frac {F} {\surd 5} - 0.5)                   \\       
\end{array}
$$

What is the number of odd divisors in a number $n$? It is at least 1 
and at most  $N = \lfloor \lg n$. It is easy to show that the number of odd divisors
 tends to  $(N+1)/2$. Thus, there are on average about 1.5 multiplications executed
 by the loop body. For all n, if k is the number of multiplications  per loop iteration,
 then  $1 \leq k \leq 2$ compared to other methods where $ k \geq 2$. 
The loop has $\lfloor \lg n +1$  iterations and {\it Golden} is $O(\lg n)$ in the 
number of loop executions.

For $n =2^{m}$, only one multiplication is required in a loop step compared to 
the two multiplications for {\it Binet} and the two multiplications for {\it Cullhow}.
There are two practical difficulties: at least full size multiplication is required in each 
loop iteration (otherwise a size adjusting process is required), and error can occur  
due to finite approximation of irrational numbers. Integer methods avoid both 
of these difficulties.

This can be converted to a simple $O(\lg n)$ recursion, with {\it Rgolden} calling 
the recursive procedure {\it Rgold}. Our recursion, compared to that for powers by 
Rawlins [7], uses one less variable and 3 less assignments by adding an {\it else}. 
A small revision gives a recursive solution for Lucas.

$$
\begin{array}{l}
 Procedure: \ Rgold(n) \ given \ n \geq 1, return \ \phi^{n}   \\

    if ( n =1) \ then \  return \ ( \phi) \\
   
if ( odd(n) )\ then \ \ return \ (\phi* Rgold(\frac {n} {2})^2) \ else \ 
return \ (Rgold(\frac {n} {2})^2)

\end{array}
$$

$$
\begin{array}{l}
 Algorithm: \ Rgolden(n) \ given \ n \geq 0, return \ F_{n}   \\
    if ( n \leq 1) \ return \ (n)    \\
    return \  \lceil( \frac {Rgold(n/2)} {\surd 5} - 0.5)        \\       
\end{array}
$$

\subsection{ Integer $O(\lg n)$ Algorithms}

In [4] {\it Cullhow}, using a product of Lucas numbers [3,5] was claimed to be the best. 
It used one square and one conventional multiplication but was only defined for $n=2^m$, 
techniques to extend to all $n$ are well known.  
In [5], the idea of using Lucas numbers was extended from $n=2^{m}$ to all $n$ and introduced 
the idea of using only two squares. This author claimed that performing squares  using 
FFT based multiplication  would result in a faster execution than the one multiplication 
one square method when  $n$ was sufficiently large. 
As can be seen in section 1.1 the resulting algorithm is rather complicated and the referred  FFT based multiplication is represented by $*$, clearly, 
the algorithm does not implement FFT based multiplication and n would need to be very large indeed to attain the theoretical improvement.
The method we use latter can also use FFT based multiplication so comparsions need only be made on the representations in this paper. 
 

We give an algorithm where the central calculation for doubling is derived as follows.
Using the well known identities [1], $ F_{n+m} = F_{m} * F_{n+1}+ F_{m-1}*F_{n}$ 
and   $ F_{n+1} * F_{n-1} =  F_{n}^2 + (-1)^n$ , it is an exercise to derive:
$$  F_{2k+1} = F_{k+1}^{2}+ F_{k}^{2} $$ 
$$  F_{2k-1} = F_{k}^{2}+ F_{k-1}^{2} $$ 
$$  F_{2k} = F_{2k+1} - F_{2k-1} $$ 

This can be executed in three squares by rearrangement. The following doubling 
calculation that uses two multiplications is more efficient and is derived as 
follows. First note that when $m=n$ an equation with two multiplications 
can be factored to give one multiplication.
 
$$  F_{n+m} = F_{n+n} = F_{2n}= F_{n} * (F_{n+1}+ F_{n-1})$$

This gives a term for $F_{2k}$   but as seen above neither  $F_{2k+1}$ nor $F_{2k-1}$ factor. Trying the 
term $F_{2n -2}$, for $m= n-2$ gives $F_{n+(n-2)}= F_{n-2} * F_{n+1}+ F_{n-3}*F_{n}$,
which does not factor and would require five adjacent terms. Consider, 
$F_{2n -2}= F_{n+n-1-1}= F_{(n-1)+(n-1))}= F_{n-1}*( F_{n} +F_{n-2})$.
 
We now have four adjacent terms that determine their doubling equations.
 $$  F_{2k-2} = F_{k-1}*( F_{k} +F_{k-2})$$
 $$  F_{2k} = F_{k}*( F_{k+1} +F_{k-1})$$   
 $$  F_{2k-1} = F_{2k} - F_{2k-2} $$ 
 $$  F_{2k+1} = F_{2k} + F_{2k-1} $$ 

This second doubling method requires four adjacent sequence terms and is not related 
to  $2 \times 2$ matrix methods, which only use three adjacent sequence terms [3,4].
Rather than use $F_{n+m}$ to update as in the $odd(i)$ step of {\it Golden}( which is complex), 
simply shift the four sequence terms one step forward as required. When to do this is 
determined by the odd divisors of $n$ in reverse. Reference to {\it Rgolden} will make this 
clear. The improvement in [5] over {\it Cullhow} was to replace the multiplications with two  
squares, assuming an efficient FFT based method of computing squares and to extend it to 
all $n$. It was estimated that $S(n) = {2 \over 3}M(n)$. This results in a tie for our two sets 
of equations.Using Occam's Razor, we claim our two multiplication method is better for computer 
implementation, where the FFT speedup is not really practical. 


\subsection{ Alternate, an integer iteration}

The logarithmic power method for all $n$ used in {\it Golden} was not efficient when 
applied to our formulas   and was replaced with one based 
on the {\it Dgolden} recursion. The doubled sequence of four adjacent terms 
is shifted up one position as necessary. 
For simplicity, the sequence variables are renamed as follows: 
 $FLL = F_{k-2}, FL = F_{k-1},FM = F_{k},   FH = F_{k+1}  $

$$
\begin{array}{l}
 Algorithm: \ Alternate(n) \ given \ n > 0, \ return \ FM\leftarrow F_{n}   \\
     N \leftarrow \lfloor \lg n  \\
    
     array  \ markOdd[N] \leftarrow 0                          \\
     i \leftarrow  n   ;     j \leftarrow N                                    \\ 
   while(j >0)\{if ( odd(i)) \ markOdd(j)= 1; i= i/2;j=j-1 \} \ \\
  
      FLL \leftarrow  1 ; FL \leftarrow  0 ; FM \leftarrow  1 ; FH \leftarrow  1 \\
     j = \leftarrow 1 \\                     
     while (j\leq N) \ \  \\
    \ \ \ \      FLL \leftarrow FL *(FM +FLL)           \\
    \ \ \ \      FM  \leftarrow        FM *(FH + FL)    \\
    \ \ \ \      FL \leftarrow  FM-FLL                 \\
    
    \ \ \ \    if ( markOdd(j)) \{FLL \leftarrow FL; FL \leftarrow FM;
                                 FM \leftarrow FL + FLL \  \} \\
\ \ \ \      FH \leftarrow  FM + FL                   \\

               \ \ \ \  j \leftarrow j+1; \ \ \	\\ 
    endwhile \\
         return \ FM    
\end{array}
$$

Moving the FH update to below the if reduces the cost of shift updating.
Two multiplications and four to five additions in the iterative loop make this very fast 
compared to [5]. 
No pre conditions nor post conditions are used, as were required in [5]. 
All general methods need some kind 
of odd(i) selected calculation, when n is not a power of 2, and this has been reduced 
here to the insertion of a shift forward of two sequence terms and one addition.  
Thus, {\it Alternate} 
is competitive with the two best to date for when squaring is faster than multiplication [5] 
for integer arithmetic, and we claim 
the algorithm to be simpler and easier to understand.

\section{Execution Analysis}

Computer execution is affected by the growth of the number of storage bits, 
$\eta_{k} $, required to represent and manipulate $F_{k}$ as $k$ approaches $n$. 
Previous papers did not consider the maximum Fibonacci number that could be 
calculated by a program. 

$$ \eta_{n} = \lfloor \lg F_{n} +1$$
$$ \eta_{n} = \lfloor \lg (\lceil ( {\phi^{n} \over \surd5} -0.5) +1$$
$$ \eta_{n} \approx  \lfloor ( n \lg \phi - 0.5 \lg 5 )+1 $$
$$ \hat n_{max} \approx  \lceil {\eta_{n} + 0.5 \lg 5 - 1  \over\lg \phi} $$  

For a given $\eta_{n}$,we can now estimate the largest index  $n$ 
for a Fibonacci number that can be represented by a data type
(and, conversely, estimate the bits required to store  $F_{n}$ for a given$n$.) 
Given 128 bit signed integer representation, the estimated maximum value would be $n = 184$.
For input $n=2^8$, a 177 bit word is estimated. 
Table 1 shows some estimates and  actual values for programs

\begin{table}[hbtf]
\caption{For signed Type size $ \eta_{n}$ Max $n$ values }
\begin{tabular}{|l|l|l|l|l|}
\hline
$ \eta_{n}$ & Type & $\hat n_{max}$ & AC & Gold\\ \hline 
\hline

24 & 32 float & 34 & na & 30\\ \hline

31  & 32 int & 45 & 46 & 36 \\ \hline
53 & 64 real &  76 & na & 74 \\ \hline

63 & 64 long &  91 
 & 92 & na \\ \hline

90,995 &  integer & n=  $2^{17}$ &  &  \\ \hline

\end{tabular}
\end{table}

The immediate objection to {\it Golden} is the use of irrational numbers and the possibility of
computer error due to truncation. Using 64 bit long  and 64 bit double in Java, {\it Golden}
failed at $F_{75}$, differing by 1 in the last digit from {\it AC}, which failed 
at $F_{93}$ from overflow. Note that a double has fewer digits than a long of the same 
computer bit size representation, so part of this failure is a practical limitation due to 
the representation of reals in the computer. In Table 1, assuming no truncation error, the 
predicted maximum with an effective  mantissa of 53 bits is at $n=76$, the actual maximum,
 including truncation error, is  at $n=74$. Assuming additional space, 
{\it Golden} works correctly for any $n$. 
 
To examine truncation error for $\phi $ and $\surd 5$, these were truncated in the programs to 
9 places, representing 31 bit integers. The {\bf truncated} {\it Golden} failed at  
$F_{37}$ differing by 1 in 
the last digit from  a 32 bit int version of {\it Alternating}, which failed at $F_{47}$ 
from overflow. For 32 bit floating point 
{\it Golden} failed at $F_{31}$ differing by 1. This truncation error explains why the 
estimates $\hat n_{max}$ for $n_{max}$ are different  than the actual values found.

A secondary objection to {\it Golden} is that full multiplication of size $\eta_{n}$ is 
required in each 
iteration  compared to the integer methods that only require size multiplication $\eta_{k}$ 
for loop iteration $k$.
Assuming that the cost of multiplying numbers with k positions is $k^2$ (the best case  
for {\it Golden} with a single multiplication), the average cost of multiplication 
for {\it Golden} is the full size cost $\overline M_{G} = \eta_{n}^{2}$.
Assuming $n= 2^k$, the average cost of multiplication for {\it Alternating} with two 
multiplications is $\overline M_{A} = { 2 \over k} \sum_{i=1}^{i = k} i^2$. 
Thus, $\overline M_{A} \approx {2 \over 3 }\overline M_{G}$ which means that {\it Alternate} 
has lower total multiplication costs than {\it Golden}, when the size of the multiplication 
is taken into account. If the cost of multiplication only depends on the register size, 
then {\it Golden} is better.

\section{Experiments}

In [4] experiments were run only on powers of two from $n= 2^{8}$ up to $n= 2^{17}$. Our 
analysis indicates that these runs were on overflow values (or used an extended representation via sofware)
and so did not measure the real register cost of computer multiplication and addition. 
Because of its simplicity, the timing of Binet should have been much faster.
There are a number of possible reasons that would explain why Binet appeared slower.
Running an overflow experiment on $n= 2^{17}$, {\it Golden} was faster than  {\it Alternate}, 
and the linear {\it Tumble} was orders slower.  

It is difficult to do timing on fast modern multitasking systems. For instance at the time of 
writing, Java  System.currentTimeMillis() is useless for these algorithms with resolution of 
20 ms (run on a sigle CPU  Windows 2000 system).  
Timing in Java proved to be problematic at best. A native timer was used that gave 
resolution to about .004 ms, when background noise was low. However, repeat times varied to 
the extent that we do not find the method  reliable for other than broad conclusions. 
Repeated runs indicated that {\it Golden} 
was maybe faster at $F_{40}$ and that iteration was maybe slower at $F_{92}$.
Although average run times of programs can be measured,   
the Java optimization methods make Java unsuitable for experimental evaluation of 
algorithms because of run optimization.

The algorithms were recoded in C and timing experiments were run. Again it was difficult 
to measure results. At $n=92$ with I64 integers, 
{\it Alternating}  was a 
bit slower than {\it Tumble}.
However, {\it Alternating} using 32 long was a bit faster. In overflow, {\it Golden}
was a bit faster than all others. For $n= 2^{10}-1$ (the floating point overflow limit 
for {\it Golden}), {\it Tumble } took somewhat longer than 
{\it Alternating}, which took about twice as long as {\it Golden}(an unfair comparsion). 
On modern systems, 
multiplication is faster than the theoretical assumptions in [4] and [5] and, of course, 
extended arithmetic has additional execution costs; therefore, experimental verification 
of theoretical results can be difficult. The more serious conclusion is that a theory not 
based on modern systems may have difficulty predicting execution performance.

\section{Conclusions}
We were able to estimate the largest Fibonacci number that can be represented in 
a given finite storage.
This was not done in previous papers. For 64 long $\hat n_{max}= 91$ closely agreed with $F_{n_{max}= 92}$ for 
integer programs. For 64 bit floating $\hat n_{max}= 76$ closely agreed with 
$F_{n_{max}= 74}$ for  {\it Golden}. The float estimate errors  are higher because of 
the truncated representation of irrationals. 

{\it Golden} has the best iteration multiplication costs of  $O(2 \lg n)$, 
$\Theta(1.5 \lg n)$ and  $\Omega(1 \lg n)$. 
{\it Golden} and   {\it Dgolden} provide the 
most simple constructive proofs that Fibonacci numbers can be found in $O( \lg n)$ time.
Their practical limitations result from errors accumulating from finite representation 
of irrationals, the requirement for floating point, and by full multiplication costs 
because of the irrational constants. However, by increasing word size appropiately, 
they can always give $F_{n}$. Even assumming the efficient squaring of [5], S(n)= 2/3M(n), 
gives a multiplication cost of 7/6 for {\it Golden} compared to 8/6 for [5].
This leads to the surprising result that, given sufficient storage, and 
assuming multiplications have storage size, {\it Golden} is the fastest.  

{\it Alternating} compared to [5] does not require the 
introduction of Lucas numbers  and is much simpler. An orginal aspect is 
the use of four sequence terms generated by only two multiplications, unlike matrix based 
equations that use only three sequence terms but require three multiplications.
It is $O(2 \lg n)$,   $\Omega(2 \lg n)$, and is as fast and more practical than [5]
for computation.

Several algorithms were encoded in Java and some execution results were obtained.
$F_{92}$ was the largest number found for each integer program, when using 64 bit Java long.
$F_{74}$ was the largest number found for {\it Golden}, when using 64 bit floating 
point. Time comparisons for {\it Golden} were limited by $n=74$, where it appeared faster 
than other methods.

The run times in [4] begin at $F_{n = 2^8}$, which requires about 178 integer bits, and are 
not valid for measuring the effect of multiplication cost as intended. Although the calculated 
values appeared in agreement with run times [4], these theoretical values were based on the run 
time of a large input, contaminating the theory results. 
Run times for our integer programs were limited by $n= 92$, when using Java 64 bit long. 
The timer could not measure any real difference, but 
the log programs were slightly faster than the linear.

With present day computers, arithmetic operations are very efficient and we may assume 
a calculation model where  arithmetic operations and assignment have equal cost. Control statements are the more complicated 
operations as found in our experiments. A simple compution model would be to count each operation (keyword) as a cost of one including 
if, then, else, while, endWhile as a one cost. In other words, a key work has a cost of one. This is a reasonable computation model 
for comparing algorithms.  

An important measure of the complexity of an algorithm is readability. Human readability is enhanced by shorter code and the reduction of  ifthenelse structures that interrupt sequential flow.  So the algorithm representation is part of the overall efficienty of an algorithm 
and reduces errors when being implemented. Our design rule is as simple as possible as complex as necessary.





\bibliographystyle{elsarticle-num}
\bibliography{<your-bib-database>}





\end{document}